\documentstyle[epsfig]{mn}

\title[Constraints on planet formation across cosmic time]{Constraints on planet formation via gravitational
  instability across cosmic time}

\author[J. L. Johnson and H. Li]{Jarrett L. Johnson\thanks{E-mail: jlj@lanl.gov} and Hui Li \\
Los Alamos National Laboratory, Los Alamos, NM  87545, USA  \\
Nuclear and Particle Physics, Astrophysics and Cosmology Group (T-2)}

\pubyear{2012}

\begin{document}
\maketitle

\begin{abstract}
We estimate the maximum temperature at which
planets can form via gravitational instability (GI) in the outskirts of early circumstellar disks.  
We show that due to the temperature floor set by the cosmic microwave
background, there is a maximum distance from their host stars beyond which
gas giants cannot form via GI, which decreases with 
their present-day age.  Furthermore, we show that planet formation via GI is not possible
  at metallicities $\la$ 10$^{-4}$ Z$_{\odot}$, due to the reduced
  cooling efficiency of low-metallicity gas.  This critical
  metallicity for planet formation via GI implies a
  minimum distance from their host stars of $\sim$ 6 AU within which planets cannot
  form via GI; at higher metallicity, this minimum distance can be
  significantly larger, out to several tens of AU.  We show that these maximum and minimum
  distances significantly constrain the number of observed planets to
  date that are likely to have formed via GI at their present
  locations. That said, the critical metallicity we find for GI is
  well below that for core accretion to operate; thus, the first
  planets may have formed via GI, although only within a narrow region
 of their host circumstellar disks.
 \end{abstract}


\section{Introduction}
When did the first planets form and what were their properties?  The 
answers to these questions depend critically on the process by which 
the first planets formed.  There are two main mechanisms of planet 
formation that are widely discussed (e.g. Papaloizou \& Terquem 2006;
Youdin \& Kenyon 2012): core accretion, in which dust 
coagulates into larger and larger bodies which become the cores of 
planets (e.g. Pollack et al. 1996); and gravitational instability
(GI), in which the self-gravity of a circumstellar disk triggers the 
fragmentation and collapse of gas into a gas giant planet directly
(e.g. Boss 1997).  

In previous work, we have discussed the formation of the earliest planets 
via core accretion (see also Shchekinov et al. 2012 for related
calculations).  In this scenario, we estimated the minimum, or
'critical', metallicity to which the gas must be enriched before
planet formation can begin (Johnson \& Li 2012; hereafter Paper I).  We found that the
critical metallicity is a function of the distance from the host star,
and that the minimum metallicity for the formation of Earth-like
planets is likely to be $\sim$ 0.1 $Z_{\odot}$.  Furthermore, we were
able to show that our prediction of the critical metallicity was
consistent with the data that were available on planetary systems, with
no planets lying in the 'Forbidden zone' in which their metallicities
imply formation times longer than the time available (i.e. the disk
lifetime).   

Recently, however, Setiawan et al. (2012) have announced the discovery of a planet 
orbiting a star with a very low iron abundance of [Fe/H] $\simeq$ -2
(HIP 11952b) which appears to lie in
the Forbidden zone for planet formation via core
accretion (if this claim proves correct; see Desidera et al. 2013 in prep).  This
implies that this metal-poor planet likely formed via some other
process, such as GI.\footnote{Alternatively, this planet may have
  initially formed closer to its host star via core
  accretion 
and then migrated outward.}  Furthermore, if it is true that planets can form
via GI at metallicities below the critical value for core accretion, then the first
planets to form may have been gas giants formed via GI.  Terrestrial
planet formation (and perhaps the emergence of life), 
which likely must instead occur via core accretion,\footnote{Gas
  giants can be formed via GI, as this process involves the
  gravitational collapse of gas fragments; terrestrial planets,
  however, are by definition not gas-dominated and so likely form via
  some other mechanism, such as core accretion (but see Boley et al. 2010; Nayakshin 2010 on how GI may yield terrestrial planets).} may
only occur at later stages of cosmic history when the process of metal
enrichment has progressed further.   This then raises the question of, 
instead of the critical metallicity for planet formation via core
accretion, what conditions must be satisfied for
the first planets to form via GI. Here we address two aspects of this 
question, namely the impacts of the cosmic microwave background (CMB)
and of the reduced cooling efficiency of low-metallicity gas in regulating
planet formation via GI.

In the next Section, we review the conditions required for the
formation of planets via GI.  In
Section 3, we consider the constraints placed on this model due to the temperature floor imposed at
high redshifts by the CMB, and in Section 4 we
estimate the minimum metallicity necessary for planet formation via GI.
In Section 5 we compare our predictions for when the first
planets form via GI to the available observational data.  
We give our conclusions in Section 6.

\section{Conditions for Planet Formation via gravitational instability}
We begin by reviewing the conditions required for GI in a circumstellar disk, which we will use to derive
constraints on where planets can form in early disks via GI.  

The first condition for fragmentation to occur in a thin disk
is that $Q$ $\la$ 1, where  (e.g. Toomre 1964; Boss
1998)\footnote{Instabilities leading to fragmentation can occur even for
  $Q$ $\simeq$ 1.4 - 1.7 (e.g. Mayer et al. 2004;
  Durisen et al. 2007), but adopting such slightly higher values would
  not affect our results strongly.}

\begin{eqnarray}
Q & = & \frac{0.936 c_{\rm s} \Omega}{\pi G \Sigma}  \nonumber \\
& \simeq & 30 \left(\frac{m_{\rm *}}{1 \, {\rm M_{\odot}}}
\right)^{\frac{1}{2}} \left(\frac{T}{100 \, {\rm K}}
\right)^{\frac{1}{2}} \nonumber \\
& \times & \left(\frac{\Sigma}{10^2 \, {\rm g} \, {\rm cm}^{-2}}
\right)^{-1}  \left(\frac{r}{10 \, {\rm AU}} \right)^{-\frac{3}{2}} \mbox{\ .}
\end{eqnarray}
Here $c_{\rm s}$ is the sound speed of a gas at temperature $T$ 
with mean molecular weight $\mu$ = 2 (close to the value 
for fully molecular gas) and adiabatic index $\gamma$ = 2, $\Omega$ is the Keplerian 
angular velocity (which here we assume is identical to the epicyclic
frequency), $\Sigma$ is the surface density of the disk, and $r$ is
the distance from the central host star.  
In the second equality we have assumed a disk temperature
normalized to $T$ = 100 K at
$r$ = 10 astronomical units (AU), and a disk surface density normalized to $\Sigma$ = 10$^2$ g
cm$^{-2}$ at $r$ = 10 AU.  Finally, we have normalized the
above formula to a central stellar mass of $m_{\rm *}$ = 1 M$_{\odot}$.
According to equation (1), fragmentation is only possible in sufficiently dense and/or cold disks, 
and/or far out from the host star, especially if it is massive.  We
shall use this condition in Section 3 to derive
constraints on planet formation via GI due to the temperature floor
set by the CMB.

The second requirement for planet formation via GI
is that a circumstellar disk also cools sufficiently fast (see e.g. Gammie 2001; Nayakshin 2006; Levin 2007).  Indeed,
it is disk cooling which is thought to drive disks to $Q$ $\simeq$ 1 (e.g. Goodman 
2003; Thompson et al. 2005).  We shall use this condition, along with
the requirement that $Q$ $\simeq$ 1, in Section 4 to explore the
limits on first planet formation via GI due to the limited cooling
efficiency of low-metallicity gas.

\section{Constraints from the CMB}
Here we estimate the
maximum temperature at which planet-mass fragments may form and we
then use this to derive the limits on planet formation via GI due to the temperature floor set by the
CMB.

\subsection{Minimum Planet Mass}
Here we define the minimum mass of a planet formed via GI,
as a function of the properties of the disk.  We follow Kratter
et al. (2010; see also Rafikov 2005; Levin 2007; Cossins et al. 2009;
Forgan \& Rice 2011) 
who estimate the initial mass of fragments formed via GI
(in a disk with $Q$ $\simeq$ 1) as that on the scale of the most unstable wavelength.
This yields the following for the minimum planet mass (in units of 
the mass M$_{\rm J}$ of Jupiter):
 
\begin{eqnarray}
m_{\rm min} & \simeq & \Sigma \left(\frac{2 \pi c_{\rm s}}{\Omega} \right)^2 \nonumber \\
               & = & 0.5 \, {\rm M}_{\rm J} \, \left(\frac{m_{\rm *}}{1 \,
     {\rm M_{\odot}}}  \right)^{-1} \left(\frac{T}{100 \,
     {\rm K}} \right) 
 \nonumber \\
& \times & \left(\frac{\Sigma}{10^2 \, 
     {\rm g} \, {\rm cm}^{-2}} \right)  \left(\frac{r}{{\rm 10 \, AU}}  \right)^{3}  \mbox{\ .}
\end{eqnarray}
Simulations of disk fragmentation via GI also suggest that this
is a reasonable estimate for the minimum mass of planets (e.g. Boley
2009; Stamatellos \& Whitworth 2009).  Indeed, given that this is the
initial mass scale of fragments, it is very likely that the
final mass they achieve via the continued accretion of gas will be
much higher than this value.  As noted by Kratter et al. (2010), if
such fragments accreted enough gas to attain their isolation mass
they will have greatly overshot the planet mass range and may end
up instead as e.g. brown dwarfs.  As noted by these authors, it appears that some mechanism must halt
accretion in order for planet-mass objects to survive (see also
e.g. D'Angelo et al. 2010; Boss 2011).  One
possibility is that the circumstellar disk is photoevaporated
(e.g. Gorti \& Hollenbach 2009; Ercolano \& Clarke 2010) or otherwise
disappears (e.g. Melis et al. 2012) before accretion to super-planet
mass scales occurs.  

As we shall show next, our adoption of the {\it minimum}
planet mass allows to estimate a {\it maximum} disk temperature at which
planets may form via GI.

\subsection{Maximum Disk Temperature for Planet Formation}
To ensure that fragments which arise in the disk are not too large
to be classified as planets we must have $m_{\rm min}$ $\la$ 13
M$_{\rm J}$, which is the commonly adopted upper mass limit for
planets -- above this mass deuterium burning occurs and we assume the
object to be a brown dwarf.  We can combine equation
(2) with $Q$ = 1 in equation (1) to obtain the maximum 
disk temperature $T_{\rm max}(r)$ from which a planet of mass 
$m_{\rm min}$ can form from fragmentation of the disk:

\begin{equation}
T_{\rm max} \simeq 100 \, {\rm K} \left(\frac{m_{\rm
      *}}{1 \, {\rm M_{\odot}}} \right)^{\frac{1}{3}} \left(\frac{m_{\rm min}}{13 \, {\rm M_{\rm J}}}
\right)^{\frac{2}{3}} \left(\frac{r}{10 \, {\rm AU}}
\right)^{-1} \mbox{\ ,}
\end{equation} 
where we have normalized to the maximum planet mass of $m_{\rm min}$ = 13 M$_{\rm J}$.
If the temperature of the disk exceeds this value, then planet
formation via GI may be impossible, either
because fragmentation is suppressed (see equation 1), or if fragmentation
occurs the fragment(s) form with super-planetary masses (becoming e.g. 
brown dwarfs instead of planets; see equation 2).
While temperatures below
this maximum value may be necessary for
planet formation (and are found in simulations including
radiative cooling; e.g. Nelson et 
al. 2000; Mej{\' i}a et al. 2005; Boley et al. 2006; Forgan et
al. 2011), they are not alone sufficient.  In addition, the surface density
of the disk must also be high enough that $Q$ $\la$ 1 (equation 1), and
in the case of non-isothermal disks the cooling criterion of Gammie (2001)
must also be satisfied.\footnote{We note that the cooling criterion
has been shown to always be met for isothermal disks and disks subjected to external irradiation (Kratter \& Murray-Clay 2011).}

\subsection{The CMB temperature floor}
It is well-known that gas cannot cool radiatively to temperatures lower than that 
of the CMB, given by $T_{\rm CMB}$ =
2.73 K (1+$z$), where $z$ is redshift.  Therefore, planet formation will
not be possible if it requires that the disk cools below $T_{\rm
  CMB}$.\footnote{The suppression of planet formation via GI
  due to background radiation fields has been confirmed by
  e.g. Stamatellos \& Whitworth (2008), Cai et al. (2008) and Forgan
  \& Rice (2013), although
  these authors did not consider the effect of the CMB in particular
  (see also Cossins et al. 2010; Rice et al. 2011; Zhu et al. 2012).} 
Following equation (3), this implies that at high redshifts
planet formation can only occur 
relatively close to the host star.  We can express this maximum radius
$r_{\rm max}$ for planet formation, as a function of the host star mass $m_{\rm *}$
and redshift $z$, by equating $T_{\rm CMB}$ to the maximum disk
temperature for planet formation given by equation (3).  This yields

\begin{equation}
r_{\rm max} \simeq 40 \, {\rm AU} \left(\frac{m_{\rm
      *}}{1 \, {\rm M_{\odot}}} \right)^{\frac{1}{3}} \left(\frac{m_{\rm min}}{13 \, {\rm M_{\rm J}}}
\right)^{\frac{2}{3}} \left(\frac{1+z}{10}
\right)^{-1} \mbox{\ .}
\end{equation}
Therefore, even for the highest mass stars that may have survived to
the present day ($\simeq$ 0.8 M$_{\odot}$),
planets can only form inside $r$ $\la$ 40 AU at $z$ $\ga$ 10, roughly
the epoch of the first metal-enriched star formation in the earliest
galaxies (e.g. Bromm \& Yoshida 2011).  Thus,
we would expect such old planets formed via GI to be found on
relatively tight orbits around old, low-mass, metal-poor stars. 
In the next Section, we estimate how tight these orbits can be,
given the limited cooling properties of metal-poor gas.

  \begin{figure}
    \includegraphics[width=3.3in]{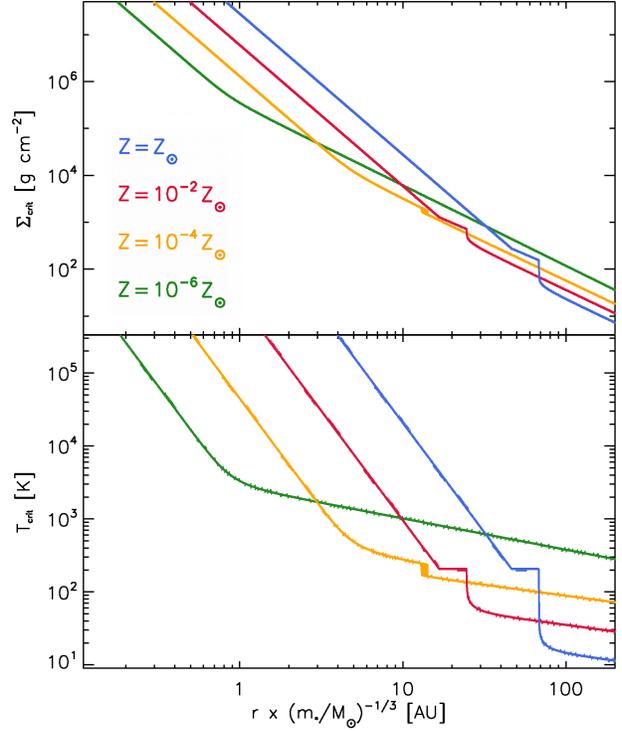}
    \caption{The critical surface density $\Sigma_{\rm crit}$ ({\it
        top panel}) and temperature $T_{\rm crit}$ ({\it bottom
        panel}) at which circumstellar disks fragment, as functions of
     distance from the host star in circumstellar disks of various metallicities, as
      labeled.  The kinks in the curves are due to breaks
      in the functional fit to the opacity, as a function of
      temperature, we have adopted from Bell \& Lin (1994). }
  \end{figure}

\section{Constraints at Low Metallicity}
To explore the effect of metallicity on the fragmentation properties of
circumstellar disks, we impose the two conditions
required for planet formation via GI described in Section 2.  We
follow the common approach of estimating the cooling rate of the disk
based on its opacity (e.g. Rafikov 2005;
Levin 2007; Kratter et al. 2010), which we assume to be proportional
to the metallicity of the disk.  Specifically, we follow exactly the
calculation presented by Levin (2007), for four different
metallicities: $Z$ = 10$^{-6}$, 10$^{-4}$, 10$^{-2}$, and 1 Z$_{\odot}$.  We
use the opacities for solar metallicity gas $\kappa($Z$_{\odot})$ given 
by equation (9) of Levin (2007; from Bell \& Lin 1994), and we scale them
with metallicity, such that $\kappa(Z)$ = $\kappa(Z_{\odot})$ $\times$ ($Z$/Z$_{\odot}$).
From this, we solve for the critical surface density $\Sigma_{\rm
  crit}$ and temperature $T_{\rm crit}$ at which the effective
viscosity of the disk reaches the critical value of $\alpha_{\rm
  crit}$ = 0.3 (Gammie 2001) and the disk fragments.  These are shown
in Figure 1, for the various metallicities we consider.  We then use
these critical values for the surface density and temperature in
equation (2) to find the minimum mass $m_{\rm min}$ of fragments
formed.  The values we find for $m_{\rm min}$, as functions of the mass $m_{\rm
  *}$ of and distance $r$
from the host star, are shown in Figure 2.

  \begin{figure}
    \includegraphics[width=3.3in]{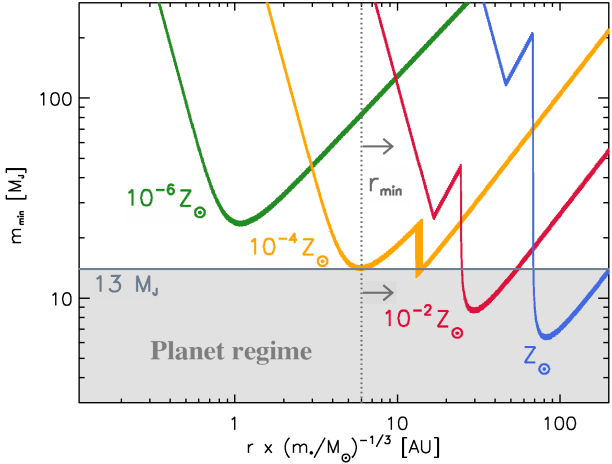}
    \caption{The minimum fragment mass $m_{\rm min}$ as a function of
     distance from the host star in circumstellar disks of various metallicities, as
      labeled.
     The critical metallicity for planet formation via GI is
      $Z_{\rm crit}$ $\simeq$ 10$^{-4}$ Z$_{\odot}$, as below this
      metallicity the $m_{\rm min}$ $\ge$ 13 M$_{\rm J}$, the maximum
      planet mass. This corresponds
    to a minimum distance from the host star of $r_{\rm min}$ $\simeq$
    6 pc (for m$_{\rm *}$ $\simeq$ 1 M$_{\odot}$), shown by the dotted
    line.  At $r$ $\le$ $r_{\rm min}$ planet
    formation via GI is not possible.  As the curves at 10$^{-2}$ and 1
    Z$_{\odot}$ show, $r_{\rm min}$ is even larger
    at higher metallicity; hence the arrows denoting the value of
    $r_{\rm min}$ shown here to be a lower limit.  Note that $m_{\rm min}$ increases at
    large radii, despite the decrease in $\Sigma_{\rm crit}$, due to
    its strong $r$-dependence via $\Omega(r)$ in equation (2).}
  \end{figure}

For the case of solar metallicity, we successfully reproduce the
results presented by Levin (2007), as expected. \footnote{But note that we have
  adopted a more widely used formula for 
  $m_{\rm min}$ (equation 2; e.g. Rafikov 2005; Cossins et al. 2009;
  Kratter 2010).  Also, we have used Kepler's Law to convert from
the rotation period of the disk, in which the results are presented by
Levin (2007), to the distance $r$ out from the central
star.}  For the lower metallicity cases, the
effect of the reduced opacity of the disk is that the critical
temperature $T_{\rm crit}$ and surface density $\Sigma_{\rm crit}$ are
significantly higher (lower) at large (small) radii, as shown in Fig. 1. 
In turn, this translates into larger minimum fragment masses
at smaller radii, at lower metallicities, as shown in Fig. 2.  
At solar metallicity, our result for the minimum fragment mass is
similar to that found by e.g. Rafikov (2005), and our result 
that fragmentation into planet-mass objects is still possible even at
10$^{-2}$ Z$_{\odot}$ is similar to that found by Meru \& Bate
(2010), assuming a metallicity dependent opacity similar to what we
have adopted (see also e.g. Boss 2002; Cai et al. 2006; Helled \& Bodenheimer 2011 on the susceptibility of
low-metallicity disks to GI) .
Our calculation is also broadly consistent with previous work which 
has shown that planet formation via GI
is difficult to achieve within tens of AU at metallicities near
the solar value (Stamatellos \& Whitworth 2008; Clarke \& Lodato 2009; Rice \& Armitage 2009; Rogers \& Wadsley 2011; Kimura \&
Tsuribe 2012; Vazan \& Helled 2012), as discussed by e.g. Boley et al. (2009) and Boss (2012),.

At metallicities $Z$ $\la$ 10$^{-4}$ Z$_{\odot}$ there exists no region of
the disk where $m_{\rm min}$ $\la$ 13 M$_{\rm J}$, the maximum 
planet mass; therefore, we interpret this to be the critical
metallicity for planet formation via GI.  
This critical metallicity for GI is well below that for
core accretion (Paper I), which implies that the first planets may
well have formed via GI in very low-metallicity circumstellar disks in
the early universe.  We note furthermore that  this critical
metallicity is below estimates of the metallicity to which the
primordial gas is enriched by the first supernovae 
($\simeq$ 10$^{-3}$ Z$_{\odot}$; e.g. Wise \& Abel 2008; Greif et
al. 2010), and this suggests that the first planets may have even
formed around second generation stars via GI.

As shown in Fig. 2, at succesively lower metallicity, the minimum fragment mass drops into
the planetary regime at smaller radii.  Thus, the critical metallicity
also implies a minimum distance $r_{\rm min}$ from their
host stars at which planets can form via GI.  From Fig. 2, we see that
at the critical metallicity of $\simeq$ 10$^{-4}$ Z$_{\odot}$, this
minimum distance is $r_{\rm min}$ $\simeq$ 6 AU, for a $m_{\rm *}$
$\simeq$ 1 M$_{\odot}$ host star.  The exact value of $r_{\rm min}$ changes slightly for
different host stellar masses, with $r_{\rm  min}$ $\propto$
m$_{\rm *}^{1/3}$ as shown on the x-axis in Fig. 2.  
At separations smaller than the $r_{\rm min}$ shown in Fig. 2, the minimum fragment mass is super-planetary,
regardless of the metallicity of the gas.  
We plot this value for
$r_{\rm min}$ in Figure 3 to show how this minimum distance and the
$r_{\rm max}$ we found in Section 3 bracket a region in which planet
formation via GI is possible.  We refer to the region inside $r_{\rm
  min}$ as the 'metal cooling-prohibited' region, as it is ultimately
the limited cooling efficiency of the gas at low metallicities that
sets the critical metallicity corresponding to $r_{\rm min}$.
From Fig. 3, we see that just $\sim$ 100 Myr after the Big
Bang the CMB temperature drops to values low enough to make $r_{\rm min}$
$\le$ $r_{\rm max}$, and so for planet formation via GI to proceed.
Thereafter, planets can form via GI in larger regions of
their host circumstellar disks at later cosmic times.

Another key point that is evident from Fig. 2 is that at
higher metallicity $r_{\rm min}$ becomes larger, further constraining
the radii in metal-enriched circumstellar disks in which planet
formation via GI is possible.  In particular, at $Z$ $\ga$ 0.1 Z$_{\odot}$
we find that $r_{\rm min}$ $\ga$ 50 AU, which is larger than $r_{\rm max}$ set
by the CMB at $\la$ 2 Gyr after the Big Bang (or at $z$ $\ga$ 3), as
shown in Fig. 3.  This suggests that planet formation
via GI may be difficult to achieve at metallicities higher than just $\sim$
10 percent of the solar value at these early times. 

While in Fig. 3 we present our results for the CMB-prohibited regime and
the metal cooling-prohibited regime independently, in principle 
they are most realistically considered together, since
at high redshift the metallicity of circumstellar disks is likely
lower and the CMB temperature is higher {\it at the same time}.
It is for simplicity that we have treated these effects 
separately, in part because the spatially inhomogeneous nature of metal enrichment implies that 
there is no clear one-to-one mapping between redshift (or $T_{\rm CMB}$) and metallicity.  That said, 
we emphasize that self-consistently including the effect of background irradiation in our calculations could 
result in somewhat larger fragments (see e.g. Levin 2007; Forgan \& Rice 2013), 
which in turn would raise the critical metallicity that we find in Fig. 2.  
As shown in Fig. 3, at early times the temperature of the CMB may 
indeed be high enough to effect such a change.


\section{Comparison with Data}
Here we compare our theoretical predictions of $r_{\rm max}$ set by
the CMB and $r_{\rm min}$ set by the critical metallicity for GI with the
star-planet separations inferred from observations.  This allows to
test whether GI is a viable explanation for the formation of the
oldest known planets.  In Fig. 3, we make this comparison,   
plotting the semimajor axes and host stellar ages of planets 
compiled in Wright et al. (2011)\footnote{We have taken the data
  directly from exoplanets.org.}
with host stars having sub-solar iron abundance ([Fe/H] $<$ 0), which we
take as an indicator of old age.  We have also  
included the four gas giants in our Solar System, as well as 
the metal-poor planetary systems reported by Sigurdsson et
al. (2003) and Setiawan et al. (2012), and the wide orbit
planets reported by Chauvin et al. (2004), Marois et al. (2008) and
Lagrange et al. (2010).
Here we have
taken the Sigurdsson et al. (2003) host stellar age to
be that of the globular cluster in which it was found, and the
semimajor axis is taken to be its original one inferred from the
modeling done by these authors.  

Also shown in Fig. 3 are the maximum radii of formation $r_{\rm
  max}$ for planets orbiting stars of three different masses: 
0.1, 0.3 and 1 M$_{\odot}$, following equation (4) with the
maximum planet mass of $m_{\rm min}$ = 13 M$_{\rm J}$.\footnote{We have
  chosen to plot the curves for this single maximal planet mass
  $m_{\rm min}$, since most of the data imply only a lower
  limit to their mass, meaning that such a high mass can not in
  general be ruled out.  We emphasize, however, that the region in
  which planet formation is suppressed is larger for
  planets with lower masses (see equation 4 and Fig. 2).}  To facilitate
the comparison with the ages of the observed planetary systems, we have
converted from redshift $z$ (in which $r_{\rm max}$ is expressed in
this equation) to the time elapsed since redshift $z$, following the
formulae describing Hubble expansion in the standard $\Lambda$CMD 
cosmology presented in e.g. Barkana \& Loeb (2001) and assuming a flat
universe with the following cosmological paramenters:  $H_{\rm
  0}$ = 70.3 km s$^{-1}$ Mpc$^{-1}$, $\Omega_{\rm \Lambda}$ = 0.73 and $\Omega_{\rm M}$ = 0.27
(Komatsu et al. 2011).  We expect the CMB temperature floor to 
suppress the formation of planet-mass objects at radii $\ga$ $r_{\rm
  max}$, in the upper shaded region of Fig. 3.  We term this the
'CMB-prohibited' zone.
  \begin{figure}
    \includegraphics[width=3.3in]{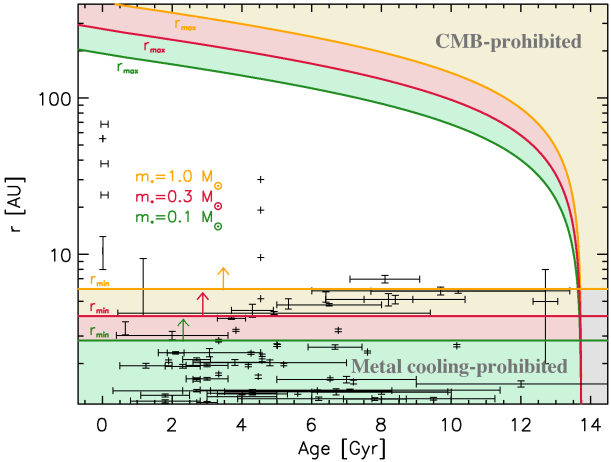}
    \caption{The semimajor axes ({\it vertical axis}) and host stellar age ({\it
        horizontal axis}) of planets, as described in Section 5, along with any reported error
      in these quantities.  The top series of colored lines show the maximum possible
      distances $r_{\rm max}$ at which planets can form from their
      host stars via GI, as a function of their present age
      (see equation 4), for four different host stellar masses as labeled.
      Beyond this maximum distance it is predicted that planet formation 
      is not possible via GI, due to the CMB temperature floor.  We
      term this region the 'CMB-prohibited' zone.  The bottom series
      of colored lines show the
      minimum possible distances $r_{\rm min}$ at which planets can
      form via GI (see Section 4), which is set by the limited cooling
      efficiency of metal-poor gas and hence defines here the 'metal
      cooling-prohibited' zone.  As shown in Fig. 2, the values of $r_{\rm
        min}$ given here are lower limits corresponding to
      metallicities just at the critical value of $Z_{\rm crit}$
      $\simeq$ 10$^{-4}$ Z$_{\odot}$; at higher metallicity $r_{\rm
        min}$ is larger.   
      Only a handful of known planets fall
      between these two zones, and so could have formed via GI at their
      present locations.}
  \end{figure}

All of the planets shown in Fig. 3 appear to lie at radii much smaller
than $r_{\rm max}$, in part because most are relatively young
(e.g. $\la$ 10 Gyr old) and formed at times when the temperature 
of the CMB temperature was low.     We also note 
that Boss (2011) argues that the formation of wide orbit gas 
giants, such as those shown at $\ga$ 20 AU, may be best explained by GI, 
especially if they are formed
around relatively massive stars, consistent with the curves in Fig. 3.

There are additional candidate planets with very wide
orbits that are not included in Fig. 1.  
These candidates, reported by Kalas et al. (2008) and Lafreni{\' e}re
et al. (2008; 2010), respectively, would lie at $\simeq$ 115 AU and $\simeq$
150 AU from their host stars, which have masses of $\simeq$ 1.9
M$_{\odot}$ and $\simeq$ 1 M$_{\odot}$,
and ages of just $\simeq$ 0.4 Gyr and $\simeq$ 5 Myr  (see also B{\' e}jar et
al. 2008; Bowler et al. 2011; and Ireland et al. 2011 for other very wide orbit $\simeq$ 14
M$_{\rm J}$ companions).  If veritable planets, they would lie just outside the 
CMB-prohibited zone and so may have formed via GI at their present locations.  Alternatively, they 
could have formed at smaller radii and migrated outward 
(e.g. Veras et al. 2009; but see Dodson-Robinson et al. 2009;
Bowler et al. 2011) or originated as free-floating planets 
(Perets \& Kouwenhoven 2012; Strigari et al. 2012).\footnote{Migration and/or capture by the host star are important caveats to consider with regard to 
conclusions drawn from comparison with data, which only reflect where
the planets orbit their host stars today.
In particular, we note that inward migration is especially likely for
planets formed via GI 
(e.g. Baruteau et al. 2011), which could potentially place some of the
planets in Fig. 1 in the CMB-prohibited zone at their formation, or
place some of those currently within $r_{\rm min}$ in between $r_{\rm
  min}$ and $r_{\rm max}$ at their formation.}  

While the planets shown in Fig. 3 lie well below the
CMB-prohibited zone, there are only a few planets that are outside the
metal cooling-prohibited zone, at $r$ $\ga$ 6 AU.  Thus, unless they
migrated inward from larger radii, it appears that there are only a
handful of known planets that could have formed via GI.
In particular, this is the case for the planets reported by Setiawan
et al. (2012).  While they 
orbit a star with [Fe/H] $\simeq$ -2, suggesting that they formed from
gas with metallicity well above the critical metallicity for GI, they
lie at $r$ $\la$ 0.81 AU, well within $r_{\rm min}$ $\simeq$ 6 AU.
Importantly, however, given the old age of $\simeq$ 12.8 Gyr inferred
for this planetary system, if
there is indeed a larger $r_{\rm min}$ of $\sim$ 25 AU for
circumstellar disks at this metallicity (and for the mass $m_{\rm *}$
$\simeq$ 0.8 M$_{\odot}$
inferred for its host star), as suggested by Fig. 2, then
this would pose a strong challenge to GI as an explanation even in this case.  

Finally, we note that it has been suggested that planets currently on 
relatively tight orbits around their host stars may have formed from the collapse
of significantly more massive (perhaps super-planetary) fragments at larger radii, which then 
migrated inward and lost mass due to tidal shear or stellar irradiation (Nayakshin 2010).
If this process is indeed at play, then it is possible that some of the planets  
in Fig. 3 may have originated from GI, despite residing in the metal cooling-prohibited
zone today.

\section{Conclusions}
As an alternative to the core accretion model for the formation of the
first planets (discussed in Paper I), we have considered
here the formation of the earliest planets via GI.  

We have argued that there is a maximum circumstellar
disk temperature only below which can planets form via GI.  In turn,
this implies a maximum distance from their host stars at
which planets can form via GI due to the temperature floor set by the
CMB.  As the CMB temperature is higher at earlier times, planets may
only form via GI at distances from their host stars which decerease
with their present-day age.  

We have furthermore estimated the minimum 
metallicity required for the fragmentation of circumstellar disks into
planet mass objects.  We find that this critical metallicity for GI is
$Z_{\rm crit}$ $\simeq$ 10$^{-4}$ Z$_{\odot}$, well below that for
core accretion.  In turn, because
planet formation via GI is possible at smaller distances from the host
star at lower metallicities, this critical metallicity implies a
minimum distance of a few AU at which planets can
form via GI.  

These two limits together imply that, while planet
formation via GI can take place at metallicities below those required
for core accretion, it can only occur at metallicities $\ga$ 0.1 Z$_{\odot}$ 
at times $\ga$ 2 Gyr after the
Big Bang.  In particular, this does not rule out that the first planets in the Universe may
indeed have formed via GI at metallicities 10$^{-4}$ $\la$ $Z$ $\la$ 10$^{-1}$ Z$_{\odot}$ 
during the epoch of the first galaxies, $\sim$ 500 Myr after the Big Bang
(e.g. Bromm \& Yoshida 2011).

That said, we find that there are only a handful of known planets which lie
within the bounds of the metal cooling- and CMB-prohibited zones
in which planets can form via GI.  It may be, however, that some known
planets could have migrated inward from their formation sites outside
the metal cooling-prohibited zone.  This may explain, in particular,
the existence of the very low-metallicity planets reported by Setiawan et
al. (2012), the formation of which is otherwise difficult to explain in the core
accretion model.

\section*{Acknowledgements}
This work was supported by the U.S. Department of Energy
through the LANL/LDRD Program.  JLJ gratefully acknowledges 
the support of a Director's Postdoctoral Fellowship at Los Alamos
National Laboratory.  The authors thank the reviewers for constructive
and cordial reports, as well as for encouraging us to explore the impact
of low metallicity on planet formation via GI as we have done in
Section 4.


\begin{thebibliography}{99}

\bibitem[2(2000)]{b}Barkana, R., Loeb, A. 2001, PhR, 349, 125
\bibitem[2(2000)]{b}Baruteau, C., et al. 2011, MNRAS, 416, 1971
\bibitem[2(2000)]{b}B{\' e}jar, V.~J.~S., et al. 2008, ApJ, 673, 185
\bibitem[2(2000)]{b}Bell, K.~R., Lin, D.~N.~C. 1994, ApJ, 427, 987
\bibitem[2(2000)]{b}Boley, A.~C., et al. 2006, ApJ, 651, 517 
\bibitem[2(2000)]{b}Boley, A.~C. 2009, ApJ, 695, L53
\bibitem[2(2000)]{b}Boley, A.~C., et al. 2010, Icar, 207, 509
\bibitem[2(2000)]{b}Boss, A.~P. 1997, Sci, 276, 1836
\bibitem[2(2000)]{b}Boss, A.~P. 1998, ApJ, 503, 923
\bibitem[2(2000)]{b}Boss, A.~P. 2002, ApJ, 567, L149
\bibitem[2(2000)]{b}Boss, A.~P. 2011, ApJ, 731, 74
\bibitem[2(2000)]{b}Boss, A.~P. 2012, MNRAS, 419, 1930
\bibitem[2(2000)]{b}Bowler, B., et al. 2011, ApJ, 743, 148
\bibitem[2(2000)]{b}Bromm, V., Yoshida, N. 2011, ARA\&A, 49, 373
\bibitem[2(2000)]{b}Cai, K., et al. 2006, ApJ, 636, L149
\bibitem[2(2000)]{b}Cai, K., et al. 2008, ApJ, 673, 1138
\bibitem[2(2000)]{b}Chauvin, G., et al. 2004, A\&A, 425, 29
\bibitem[2(2000)]{b}Clark, P.~C., et al. 2011, Science, 331, 1040
\bibitem[2(2000)]{b}Clark, P.~C., Glover, S.~C.~O., Klessen, R.~S. 2008, ApJ, 672, 757
\bibitem[2(2000)]{b}Clarke, C.~J., Lodato, G. 2009, MNRAS, 398, L6
\bibitem[2(2000)]{b}Cossins, P., Lodato, G., Clarke, C. 2009, MNRAS, 393, 1157
\bibitem[2(2000)]{b}Cossins, P., Lodato, G., Clarke, C. 2010, MNRAS, 401, 2587
\bibitem[2(2000)]{b}D'Angelo, et al. 2010, arXiv:1006.5486
\bibitem[2(2000)]{b}Desidera, S., et al. 2013, A\&A, in prep 
\bibitem[2(2000)]{b}Dodson-Robinson, S.~E., et al. 2009, ApJ, 707, 79
\bibitem[2(2000)]{b}Durisen, R.~H., et al. 2007, PrPl, 951, 607
\bibitem[2(2000)]{b}Ercolano, B., Clarke, C.~J. 2010, MNRAS, 402, 2735
\bibitem[2(2000)]{b}Forgan, D., et al. 2011, MNRAS, 410, 994
\bibitem[2(2000)]{b}Forgan, D., Rice, K. 2011, MNRAS, 417, 1928
\bibitem[2(2000)]{b}Forgan, D., Rice, K. 2013, MNRAS, accepted (arXiv:1301.1151)
\bibitem[2(2000)]{b}Gammie, C.~F. 2001, ApJ, 553, 174
\bibitem[2(2000)]{b}Goodman, J. 2003, MNRAS, 339, 937
\bibitem[2(2000)]{b}Gorti, U., Hollenbach, D. 2009, ApJ, 690, 1539
\bibitem[2(2000)]{b}Greif, T.~H., et al. 2010, ApJ, 716, 510
\bibitem[2(2000)]{b}Helled, R., Bodenheimer, P. 2011, Icar, 211, 939
\bibitem[2(2000)]{b}Ireland, M.~J., et al. 2011, ApJ, 726, 113
\bibitem[2(2000)]{b}Johnson, J.~L., Li, H. 2012, ApJ, 751, 81
\bibitem[2(2000)]{b}Kalas, P., et al. 2008, Sci, 322, 1345
\bibitem[2(2000)]{b}Kimura, S.~S. Tsuribe, T. 2012, ApJ, submitted (arXiv:1205.3013)
\bibitem[2(2000)]{b}Komatsu, E., et al. 2011, ApJS, 192, 18
\bibitem[2(2000)]{b}Kratter, K.~M., et al. 2010, ApJ, 710, 1375
\bibitem[2(2000)]{b}Kratter, K.~M., Murray-Clay, R.~A. 2011, ApJ, 740, 1
\bibitem[2(2000)]{b}Lafreni{\' e}re, D., et al. 2008, ApJ, 689, 153L
\bibitem[2(2000)]{b}Lafreni{\' e}re, D., et al. 2010, ApJ, 719, 497
\bibitem[2(2000)]{b}Lagrange, A.-M., et al. 2010, Sci, 329, 57
\bibitem[2(2000)]{b}Levin, Y. 2007, MNRAS, 374, 515
\bibitem[2(2000)]{b}Marois, C., et al. 2008, Sci, 322, 1348
\bibitem[2(2000)]{b}Mayer, L., et al. 2004, ASPC, 321, 290
\bibitem[2(2000)]{b}Mej{\' i}a, A.~C., et al. 2005, ApJ, 619, 1098
\bibitem[2(2000)]{b}Melis, C., et al. 2012, Nat, in press (arXiv:1207.1162)
\bibitem[2(2000)]{b}Meru, F., Bate, M.~R. 2010, MNRAS, 406, 2279
\bibitem[2(2000)]{b}Nayakshin, S. 2006, MNRAS, 372, 143
\bibitem[2(2000)]{b}Nayakshin, S. 2010, MNRAS, 408, L36
\bibitem[2(2000)]{b}Nelson, A.~F., et al. 2000, ApJ, 529, 357
\bibitem[2(2000)]{b}Papaloizou, J.~C.~B., Terquem, C. 2006, RPPh, 69, 119
\bibitem[2(2000)]{b}Perets, H.~B., Kouwenhoven, M.~B.~N. 2012, ApJ, submitted (arXiv:1202.2362)
\bibitem[2(2000)]{b}Pollack, J.~B., et al. 1996, Icar, 124, 62
\bibitem[2(2000)]{b}Rafikov, R.~R. 2005, ApJ, 621, 69
\bibitem[2(2000)]{b}Rice, W.~K.~M., Arimtage, P.~J. 2009, MNRAS, 396, 2228
\bibitem[2(2000)]{b}Rice, W.~K.~M., et al. 2011, MNRAS, 418, 1356
\bibitem[2(2000)]{b}Rogers, P.~D., Wadsley, J. 2011, MNRAS, 414, 913
\bibitem[2(2000)]{b}Setiawan, J., et al. 2012, A\&A, 540, A141
\bibitem[2(2000)]{} Shchekinov, et al. 2012, arXiv:1212.0519
\bibitem[2(2000)]{b}Sigurdsson, S., et al. 2003, Sci, 301, 193
\bibitem[2(2000)]{b}Stamatellos, D., Whitworth, A.~P. 2008, A\&A, 480, 879
\bibitem[2(2000)]{b}Stamatellos, D., Whitworth, A.~P. 2009, MNRAS, 392, 413
\bibitem[2(2000)]{b}Strigari, L.~E., et al. 2012, MNRAS, 423, 1856
\bibitem[2(2000)]{b}Thompson, T.~A., Quataert, E., Murray, N. 2005, ApJ, 630, 167
\bibitem[2(2000)]{b}Toomre, A., 1964, ApJ, 139, 1217
\bibitem[2(2000)]{b}Vazan, A., Helled, R. 2012, ApJ, accepted (arXiv:1206.5887)
\bibitem[2(2000)]{b}Veras, D., Crepp, J.~R., Ford, E.~B. 2009, ApJ, 696, 1600
\bibitem[2(2000)]{b}Wise, J.~H., Abel, T. 2008, ApJ, 685, 40
\bibitem[2(2000)]{b}Wright, J.~T., et al. 2011, PASP, accepted (arXiv:1012.5676)
\bibitem[2(2000)]{b}Youdin, A.~N., Kenyon, S.~J. 2012, arXiv:1206.0738
\bibitem[2(2000)]{b}Zhu, Z., et al. 2012, ApJ, 746, 110


\end{thebibliography}
\end{document}